\documentclass[aps,preprint,floats,epsf,epsfig,nofootinbib,letter]{revtex4}

\usepackage{graphicx}
\usepackage{dcolumn}
\usepackage{bm}

\def\be{\begin{eqnarray}}
\def\en{\end{eqnarray}}
\def\non{\nonumber}
\def\la{\langle}
\def\ra{\rangle}

\def\CP{{\it CP}~}
\def\cp{{\it CP}}
\def\dL{{\delta {\cal L}}}
\begin{document}

\renewcommand{\baselinestretch}{1.10}


\vskip 1.5 cm

\centerline{\Large\bf Remarks on Strong {\it CP}-Violating Lagrangians}

\bigskip
\centerline{\bf Hai-Yang Cheng$^{1,2,3}$}
\medskip
\centerline{$^1$ Institute of Physics, Academia Sinica}
\centerline{Taipei, Taiwan 115, Republic of China}
\medskip
\centerline{$^2$ Physics Department, Brookhaven National
Laboratory} \centerline{Upton, New York 11973}
\medskip
\medskip
\centerline{$^3$ C.N. Yang Institute for Theoretical Physics,
State University of New York} \centerline{Stony Brook, New York
11794}
\medskip

\centerline{\bf Abstract}
\bigskip
\small
Owing to a different treatment of the vacuum alignment,
the strong \cp-violating Lagrangian obtained by Di Vecchia, Veneziano and Witten (DVW) 3 decades ago do not look quite the same as the one originally derived by Baluni at the quark or hadron level. We show that they are consistent with each other and emphasize that,  within the DVW approach, the $\theta G\tilde G$ term is not entirely removed away after the vacuum is rotated from the \cp-odd state to the \cp-even one; strong \CP violation resides not only in the quark mass terms but also in the residual topological sector.
Contrary to some claims, it is necessary to include the SU(3)-singlet $\eta_0$ tadpole contribution for strong \cp-odd effects induced by the Baluni-type Lagrangian to ensure that strong \CP violation vanishes in the zero axial anomaly limit.

\vskip 0.3in
\noindent PACS: 11.30.Er, 11.30.Rd, 12.39.Fe

\maketitle

%
%

{\bf 1.} The commonly used strong \cp-violating quark operator first derived by Baluni \cite{Baluni} and the chiral Lagrangian derived by Di Vecchia, Veneziano \cite{DiVecchia} and by Witten  \cite{Witten} (DVW) 3 decades ago do not look quite the same. Especially, the treatment of the vacuum alignment seems to be quite different. Whether these \cp-odd Lagrangians are equivalent is the issue to be explored in this short note.

It is well known that the nontrivial topological structure of the $\theta$ vacuum in QCD not only allows for instanton solutions but also induces an additional $T$- and $P$-violating $\theta$ term to the QCD Lagrangian
\be
{\cal L}_{\rm QCD}=-{1\over 4}GG+\sum_i\bar q_i(iD\!\!\!\!/\ -m_i)q_i-{g^2\over 32\pi^2}\theta_{\rm QCD}G\tilde G.
\en
The $\theta$ vacuum is generally  $P$ and \CP noninvariant. In practice, it is more convenient to work with the \cp-invariant vacuum so that the $\theta_{\rm QCD} G\tilde G$ term is represented as an operator perturbation and \CP is explicitly broken.
This can be achieved by going to a basis where $\theta_{\rm QCD} G\tilde G$ is rotated away and replaced by an effective {\it CP}-odd operator in terms of quark fields. Owing to the axial anomaly, a chiral rotation of the quark field $q\to {\rm exp}(i\alpha\gamma_5)q$ will induce a change of $\theta_{\rm QCD}$, namely, $\theta_{\rm QCD}\to \theta_{\rm QCD}-2\alpha$.

In general, the vacuum expectation value (VEV) of the quark condensate has the expression
\be \label{eq:qcVEV}
\la \bar q_{iL}q_{jR} \ra=-\Lambda^3\delta_{ij}e^{-i\phi_i}.
\en
$P$ and \CP symmetries require that $\phi_i=0,\pi$. We now make a chiral rotation of the quark field $q_{iR}\to {\rm exp}(i\phi_i/2)q_{iR}$ so that $\la \bar q_{iL}q_{jR}\ra=-\Lambda^3\delta_{ij}$. The QCD Lagrangian then reads
\be \label{eq:LQCD}
{\cal L}_{\rm QCD}=-{1\over 4}GG+\sum_i\bar q_iiD\!\!\!\!/\ q_i-\sum_im_i\cos\phi_i \bar q_iq_i+\sum_i m_i\sin\phi_i\bar q_ii\gamma_5 q_i-(\theta-\sum\phi_i)q(x),
\en
with $q(x)\equiv (g^2/32\pi^2)G\tilde G$, where for simplicity we have dropped the subscript ``QCD" of $\theta$.
To avoid the vacuum instability, the phases $\phi_i$ should satisfy the constraints \cite{Crewther,Crewther1980}
\be \label{eq:constraints}
m_i\sin\phi_i=m_j\sin\phi_j, \qquad \sum_i\phi_i=\theta.
\en
As we shall see below, the first constraint arises from the vacuum alignment, while the second one comes from the anomalous Ward-Takahashi (WT) identity. Eqs. (\ref{eq:LQCD}) and (\ref{eq:constraints}) lead to the well known strong {\it CP}-violating operator first derived by Baluni \cite{Baluni}:
\be \label{eq:CPLquark}
\delta{\cal L}_{CP}^{\rm Baluni}=\theta\bar m(\bar ui\gamma_5 u+\bar di\gamma_5 d+\bar si\gamma_5 s), \quad \bar m=(1/m_u+1/m_d+1/m_s)
\en
for $\theta\ll 1$ and three light quarks. In principle, one can apply current algebra to compute the hadronic matrix elements induced by $\dL_{CP}^{\rm Baluni}$. \footnote{In lattice or quark model calculations, it is the disconnected insertion of $\delta{\cal L}_{CP}^{\rm Baluni}$ that is related to the insertion of $\theta G\tilde G$ in the hadronic process \cite{Aoki90}. The connected insertion of $\delta{\cal L}_{CP}^{\rm Baluni}$ must vanish for on-shell amplitudes.  Since a quark loop with the insertion of $\delta{\cal L}_{CP}^{\rm Baluni}$ at zero momentum is the same as an insertion of $\theta G\tilde G$, the trick of rotating the $\theta$ term into a pseudoscalar density does not change the nature of the calculation which must be done in the lattice or quark model framework \cite{Aoki90}. Hence, if care is not taken, the use of Baluni's Lagrangian may lead to fake results, for example, the electric dipole moment of the constituent quark \cite{Abada}.
}
However, it is much more convenient in practice to use $U(3)$ chiral perturbation theory valid in the large $N_c$ limit to get the chiral representation for the pseudoscalar quark density:
\be
\bar ui\gamma_5 u+\bar di\gamma_5 d+\bar si\gamma_5 s=-{i\over 4}f_\pi^2 v\,{\rm Tr}(U-U^\dagger).
\en
with $U={\rm exp}(2i\phi/f_\pi)$, $\phi=\sum_{a=0}^8\phi^a\lambda^a/\sqrt{2}$, Tr$(\lambda^a\lambda^b)=2\delta^{ab}$, $f_\pi=132$ MeV, and
\be \label{eq:v}
v=-2{\la \bar qq\ra\over f_\pi^2}
\en
characterizing the spontaneous breaking of chiral symmetry.
Consequently, the chiral realization of $\delta{\cal L}_{CP}^{\rm Baluni}$ in the meson sector is
\be \label{eq:LCPM}
\delta{\cal L}_{CP}^M &=& -{i\over 4}\theta\bar mf_\pi^2 v\,{\rm Tr}(U-U^\dagger)= \sqrt{3}\theta\bar m f_\pi v\eta_0+O(\phi^3)+\cdots.
\en

Using the vacuum expectation value
\be \label{eq:qVEV}
\la q(x)\ra={1\over 8}{af_\pi^2\over N_c}(\theta-\sum_i \phi_i)
\en
derived from the chiral Lagrangian approach (see below) with the parameter $a$ denoting the mass squared of $\eta_0$ in the chiral limit, Di Vecchia and Veneziano \cite{DiVecchia} and Witten \cite{Witten} obtained
a different energy minimizing condition
\be \label{eq:DVmin}
m_i\sin\phi_i={a\over 2vN_c}(\theta-\sum_i\phi_i).
\en
Hence, in QCD language one has (see Eq. (A.14) of \cite{DiVecchia})
\be  \label{eq:LDV}
{\cal L}_{\rm QCD}=-{1\over 4}GG+\sum_i\bar q_iiD\!\!\!\!/\ q_i-\sum_im_i\cos\phi_i \bar q_iq_i-(\theta-\sum_i\phi_i)\left(q(x)-{a\over 2vN_c }\sum_j\bar q_ji\gamma_5 q_j\right).
\en
The chiral realization of the last term is
\be \label{eq:LDVW}
\delta {\cal L}_{CP}^{\rm DVW}=-i{af_\pi^2\over 8N_c}\bar\theta\left( {\rm Tr}(U-U^\dagger)-{\rm Tr}({\rm ln}U/U^\dagger)\right)
\en
valid to the leading order of $\bar\theta\equiv \theta-\sum\phi_i$.
As stressed by Di Vecchia, Veneziano and Witten, an important feature of this {\it CP}-violating interaction is that it does not contain terms linear in any of the pseudoscalar fields including the flavor-singlet $\eta_0$.

A comparison of the last term of Eq. (\ref{eq:LDV})  with $\delta {\cal L}_{CP}^{\rm Baluni}$  or $\delta {\cal L}_{CP}^{\rm DVW}$ with $\dL_{CP}^M$ reveals that they are not obviously equivalent.  First, it is evident from Eq.  (\ref{eq:LDV}) that, at the quark level, if the $\theta q(x)$ term is rotated away with $\sum\phi_i=\theta$, there will be no strong \CP violation at all: A disappearance of strong \CP nonconservation in the topological gauge sector will not resurrect in the quark mass sector. Second, while the $\eta_0$ is eliminated from $\delta {\cal L}_{CP}^{\rm DVW}$, it is not so in $\delta {\cal L}_{CP}^M$.  Indeed, it has been pointed out in \cite{Cheng91} and will be stressed again below with an explicit example that for strong \cp-violating effects induced by $\delta {\cal L}_{CP}^M$, it is necessary to include the $\eta_0$ tadpole contribution to ensure that the physical effect vanishes in the zero anomaly limit. Since the \cp-odd Lagrangian $\delta {\cal L}_{CP}^{\rm DVW}$ has been popularly used in the literature (see e.g. \cite{Pich}), the purpose of this short note is to clarify the aforementioned issues.

\vskip 0.3cm {\bf 2.} Following Crewther \cite{Crewther,Crewther1980} we begin with a re-derivation  of the first constriant in Eq. (\ref{eq:constraints}).
It is known that the explicit chiral symmetry breaking terms by quark masses provide a preferred direction for spontaneous chiral symmetry breaking. When explicit and spontaneous symmetry breaking are badly misaligned, explicit symmetry breaking cannot be treated as a small perturbation (for a recent discussion, see \cite{Mereghetti:2010tp}).
To find this direction one can apply the Dashen's theorem \cite{Dashen} which states that the VEV
\be
\la U(\vec{\omega})(-{\cal L }_{\rm mass})\,U(\vec{\omega})^{-1}\ra
\en
has a minimum at $\vec{\omega}=0$, where $U(\vec{\omega})={\rm exp}(i\vec{\omega}\cdot \vec{F})$ with $\vec{F}$ being the generators of $SU(n)\times SU(n)$ and $\vec{\omega}\ll1$.
It suffices to consider the right-handed $SU(n)$ rotations which amount to replacing the phases $\phi_i$ by $\phi_i+\omega_i$, subject to the $SU(n)$ constraint $\sum_{i=1}^n \omega_i=0$.
Hence,
\be
\la -{\cal L}_{\rm mass}\ra_{\vec{\omega}}=-{1\over 2}{ f_\pi^2v}\sum_i m_i \cos(\phi_i+\omega_i)
\en
should have a local minimum at $\vec{\omega}=0$ where use of $\la \bar qi\gamma_5 q\ra=0$ has been made. Note the term $(\theta-\sum \phi_i)q(x)$ in Eq. (\ref{eq:LQCD}) is irrelevant for our purpose because of the flavor-blind nature of $q(x)$ and the constraint $\sum \omega_i=0$. As pointed out by Crewther \cite{Crewther}, instead of using the Lagrangian multiplier method elucidated by Nuyts \cite{Nuyts}, it is sufficient to consider the variation $\vec{\omega}$ as the linear combination of the variations
\be
\omega_i=-\omega_j=\omega, \qquad \omega_k=0,~~~(k\neq i,j).
\en
It follows from ${d\la -{\cal L}_{\rm mass}\ra_{\vec{\omega}}/ d\omega}=0$ that
\be  \label{eq:miphii}
m_i\sin\phi_i=m_j\sin\phi_j=\lambda+{\cal O}(\epsilon^2)
\en
is independent of the flavor index $i$ and valid to the first order in chiral symmetry breaking (usually denoted by the parameter $\epsilon\propto m_q/m_N$ which approaches to zero in the chiral limit). A second constraint comes from the anomalous WT identity which leads to
\be \label{eq:AWT}
\sum_{i=1}^n {\partial\phi_i\over\partial \theta}-1={\cal O}(\epsilon).
\en
For the explicit expression of the ${\cal O}(\epsilon)$ terms on the right hand side of the above equation, see \cite{Crewther,Crewther1980} for details.
This leads to
\be \label{eq:sumphi}
\sum_i\phi_i=\theta+{\cal O}(\epsilon),
\en
which is the second constraint in Eq. (\ref{eq:constraints})
where the integration constant is fixed by requiring $\phi_i=0$ to correspond to $\theta=0$. For small $\theta$ and $\phi_i$, it follows from Eq. (\ref{eq:constraints}) that
\be \label{eq:lambda}
\lambda=\bar m\theta +{\cal O}(\epsilon^2,\theta^2).
\en

A different minimization procedure was considered by Di Vecchia, Veneziano and Witten. The vacuum expectation value of the potential energy corresponding to ${\cal L}_{\rm QCD}$ in Eq. (\ref{eq:LQCD}) is given by
\be
\la V(\phi_i)\ra=-{1\over 2}f_\pi^2 v\sum_i m_i\cos\phi_i+{1\over 8}{af_\pi^2\over N_c}(\theta-\sum_i\phi_i)^2,
\en
where we have used Eqs. (\ref{eq:v}) and (\ref{eq:qVEV}). The minimization condition $\partial\la V(\phi_i)\ra/\partial\phi_i=0$ leads to Eq. (\ref{eq:DVmin}).

The two approaches for the vacuum alignment look quite different. The first approach due to Baluni \cite{Baluni} and Crewther \cite{Crewther} relies on the Dashen's theorem where $q(x)$ plays no role,
while the second one requires the external information on the $\phi_i$ dependence of $q(x)$ which can be obtained from the chiral Lagrangian approach. Since Eq. (\ref{eq:DVmin}) does not allow the solution $\sum\phi_i=\theta$ for $\theta\neq 0$ and $\phi_i\neq 0$, it appears naively that this is not consistent with the second constraint of (\ref{eq:constraints}). However,  a non-vanishing $\bar\theta=\theta-\sum \phi_i$  of order $\epsilon$ is allowed by Eq. (\ref{eq:sumphi}). For $\phi_i\ll 1$, it follows from
Eq. (\ref{eq:DVmin}) that
\be \label{eq:relation}
{a\over N_c}\bar\theta=\theta\left(\sum_i{1\over  2m_i v}+{N_c\over a}\right)^{-1}\approx 2\bar mv\theta,
\en
where we have neglected the term $N_c/a$ compared to $1/(2m_i v)$ as $a\approx 0.73\,{\rm GeV}^2$ and $v\approx 1.6$ GeV numerically. Therefore, the minimization condition (\ref{eq:DVmin}) obtained by Di Vecchia, Veneziano and Witten is equivalent to that of Eqs. (\ref{eq:miphii}) and (\ref{eq:lambda}). We find numerically $\bar\theta\approx 2\times 10^{-2}\theta$. This means $\sum_i\phi_i$ is indeed very close to but not exactly identical to $\theta$. The last term in the QCD Lagrangian (\ref{eq:LDV}) is identical to $\dL_{CP}^{\rm Baluni}$ except that there is a residual contribution from the topological charge density $q(x)$ owing to the fact that $\bar\theta\neq 0$. \footnote{In principle, one can make a chiral rotation of the quark field $q_i\to {\rm exp}(i\phi'_i\gamma_5/2)q_i$ to remove away the $\theta G\tilde{G}$ term completely from the QCD Lagrangian provided that $\sum_i\phi'_i=\theta$. However, a small deviation of $\phi'_i$ from $\phi_i$, the phase of the quark condensate [see Eq. (\ref{eq:qcVEV})], implies that the vacuum is not \cp-even. As stressed in the beginning, we prefer to work with the \cp-invariant vacuum so that the bulk of the $\theta_{\rm QCD} G\tilde G$ term is represented as an operator perturbation and \CP is explicitly broken.
}
At the hadron level we can use the chiral Lagrangian approach to determine the chiral realization of the $G\tilde G$ term. Given the flavor-blind nature of $q(x)$, it is natural to argue that the flavor-singlet $\eta_0$ should be one of the interpolating fields for $G\tilde G$. For the case of $\sum_i\phi_i=\theta$, there is no strong \CP violation. In this case, possible solutions to Eq. (\ref{eq:DVmin}) such as $\theta=0$ and all $\phi_i=0$ have been discussed in \cite{DiVecchia,Witten}.

In short, a major difference between the two approaches is that the sum of the phases of the quark condensate $\sum\phi_i$ is taken to be the same as the $\theta$ parameter  in the Baluni approach,  while there is a small and calculable deviation of $\sum\phi_i$ from $\theta$  in the DVW scenario. As a consequence,  the $\theta G\tilde G$ term in the DVW approach is not entirely removed away after the vacuum is rotated from the \cp-odd state to the \cp-even one; strong \CP violation resides not only in the quark mass terms but also in the residual topological sector.

\vskip 0.3cm {\bf 3.} In the chiral-Lagrangian approach, the effective meson Lagrangian respected the anomalous WT identity in the leading $1/N_c$ expansion has the form \cite{Rosenzweig,DiVecchia,Ohta,Nath}
\be
{\cal L}_M &=& {f_\pi^2\over 8}[{\rm Tr}(\partial_\mu U\partial^\mu U^\dagger)+2v\,{\rm Tr}(mU^\dagger+mU)]+{N_c\over 2}\,{1\over af_\pi^2}(\partial^\mu K_\mu)^2 \non \\
&&+{i\over 4}(\partial ^\mu K_\mu)({\rm ln\,det}U-{\rm ln\,det}U^\dagger)-\theta{g^2\over 32\pi^2}G_{\mu\nu}^a\tilde{G}^{a\mu\nu},
\en
where $K_\mu$ is the Chern-Simons current with the divergence $\partial^\mu K_\mu=(g^2/16\pi^2)G\tilde G$ and $m$ is a diagonal quark mass matrix.
Since $UU^\dagger=1$, the vacuum expectation value of $U(x)$, $\la U\ra$, can be written as a diagonal matrix $V$ with the matrix elements $(e^{-i\phi_u},e^{-i\phi_d},e^{-i\phi_s})$. Making a chiral rotation $U\to UV$ so that $\la U\ra=1$ and
\be \label{eq:L}
{\cal L}_M &=& {f_\pi^2\over 8}[{\rm Tr}(\partial_\mu U\partial^\mu U^\dagger)+2v\,{\rm Tr}(mV^\dagger U^\dagger+mUV)]+{N_c\over 2}\,{1\over af_\pi^2}(\partial^\mu K_\mu)^2 \non \\
&&+{i\over 4}(\partial ^\mu K_\mu)({\rm ln\,det}U-{\rm ln\,det}U^\dagger)-{1\over 2}\bar\theta\partial^\mu K_\mu.
\en
The use of the equation of motion yields
\be \label{eq:partialK}
\partial^\mu K_\mu=-{i\over 4}{af_\pi^2\over N_c}({\rm ln\,det}U-{\rm ln\,det}U^\dagger)+{1\over 2}{af_\pi^2\over N_c}\bar\theta=\sqrt{3}{af_\pi\over N_c}\eta_0+{1\over 2}{af_\pi^2\over N_c}\bar\theta.
\en
This gives the chiral realization of the ${1\over 2}q(x)$ term.
Putting this back to ${\cal L}_M$ gives an additional mass term
$-{1\over 2}(3a/N_c)\eta_0^2$ for the $\eta_0$ due to the axial anomaly.
The VEV of $q(x)$ as shown in Eq. (\ref{eq:qVEV}) also follows from the above equation.

It is straightforward to show that the use of the Dashen's theorem for the vacuum alignment will lead to the strong \cp-violating operator $\dL_{CP}^M$, while the minimization condition $\la \partial \la V(\phi_i)\ra/\partial \phi_i=0$ yields
$\dL_{CP}^{\rm DVW}$. Applying the relation $a\bar\theta/N_c=2\bar mv\theta$ again, we see that
$\dL_{CP}^{\rm DVW}$ is the same as $\dL_{CP}^M$ except for an additional ${\rm Tr}({\rm ln}U/U^\dagger)$ term
\be \label{eq:L'}
\dL_{CP}^{\rm DVW}\to \delta{\cal L'}_{CP}^M=-{i\over 4}\theta\bar m f_\pi^2v\left( {\rm Tr}(U-U^\dagger)-{\rm Tr}({\rm ln}U/U^\dagger)\right).
\en
This extra term comes from the residual $G\tilde G$ sector and is governed by the $\eta_0$ field. Because of the large mass of the $\eta_0$, its tadpole contribution is very small in practice and hence can be neglected. In this sense, the Lagrangians $\dL_{CP}^{\rm DVW}$ and $\dL_{CP}^M$ are equivalent. However, we would like to make a caveat here that this \CP-odd operator $\delta{\cal L'}_{CP}^M$ is not the right one for studying the zero axial anomaly behavior. This will be explained below.

\vskip 0.3cm {\bf 4.}
A common feature of the strong \cp-odd Lagrangians $\dL_{CP}^M$ and $\dL_{CP}^{\rm DVW}$ is that they cannot create Goldstone bosons from the vacuum, $\la 0|\dL_{CP}|G^a\ra=0$ with $a=1,\cdots,8$ \cite{Dashen}. \footnote{This condition can be used as the starting point for deriving the Baluni Lagrangian, see \cite{CDVW}.}
This is understandable because if Goldstone bosons can be created from $\dL_{CP}$, they will be produced so abundantly to bring a shift of the vacuum and cause a vacuum instability. However, the \cp-odd operator $\dL_{CP}^{\rm DVW}$ goes one step further by having $\la 0|\dL_{CP}^{\rm DVW}|\eta_0\ra=0$ even though the $\eta_0$ is not a Goldstone boson. As explained before, the decoupling of the $\eta_0$ from  $\dL_{CP}^{\rm DVW}$ is due to the residual $q(x)$ term which compensates precisely the $\eta_0$ field occurring in the \cp-violating quark mass terms.

\cp-violating effects induced by the QCD $\theta$ term must vanish in the chiral limit and/or in the absence of the axial anomaly \cite{Aoki}. Although the absence of strong \CP violation in the zero anomaly limit is not obviously manifest in the strong \cp-violating operator $\dL_{CP}^{\rm Baluin}$ or $\dL_{CP}^M$, a correct evaluation of strong \CP phenomena must respect the aforementioned constraints. By contrast, the advantage of $\dL_{CP}^{\rm DVW}$ is that its induced matrix element vanishes obviously in the zero anomaly limit, though not manifestly in the chiral limit.
In the following we take the \cp-violating pion-nucleon coupling $\bar g_{\pi NN}$ induced by $\dL_{CP}^M$ as an example to point out that the tadpole contribution due to the $\eta_0$ must be taken into account in order to produce the correct $a\to 0$ behavior \cite{Cheng91}.

\begin{figure}[t]
\begin{center}
\includegraphics[width=0.60\textwidth]{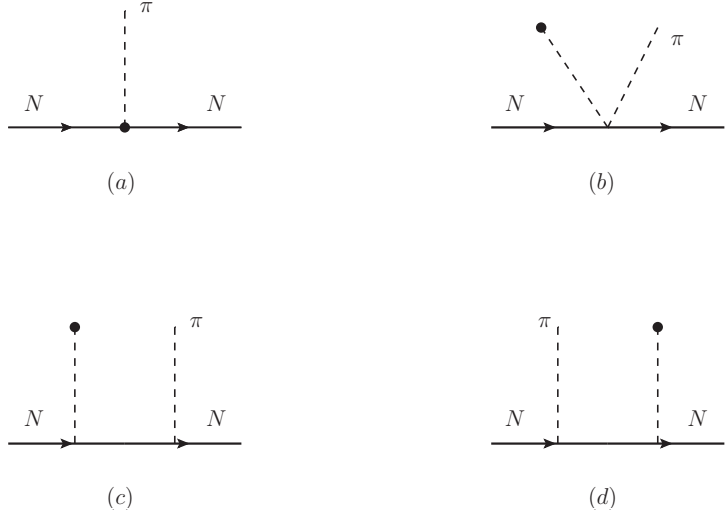}
\vspace{0.0cm}
\caption{Diagrams contributing to the \cp-violating pion-nucleon coupling $\bar g_{\pi NN}$. The dark blob indicates a vertex induced by the \cp-odd Lagrangian $\dL_{CP}^M$ or $\dL_{CP}^B$.
} \label{fig:piNN} \end{center}
\end{figure}
In the following we recapitulate the calculations in \cite{Cheng91}.
There are four diagrams contributing to $\bar g_{\pi NN}$ (Fig. 1). The strong \cp-violating operator in the baryon sector \cite{Cheng91,Pich,Cho}
\be
\dL_{CP}^B=-2\theta{\bar m\over f_\pi}[b {\rm Tr}(\bar BB\phi)+c{\rm Tr}(\bar B\phi B)],
\en
with
\be
b={2(m_\Sigma-m_N)\over 2m_s-m_u-m_d}, \qquad
c=-{2(m_\Xi-m_\Sigma)\over 2m_s-m_u-m_d}
\en
contributes to Fig. 1(a).
Fig. 1(b) arises from the $\eta_0$ tadpole contribution. Figs. 1(c) and 1(d) do not contribute to $\bar g_{\pi NN}$ owing to the derivative pion-nucleon coupling. The result is \cite{Cheng91}
\be \label{eq:gpiNN}
\bar g_{\pi NN}=2\sqrt{2}\,\theta{\bar m\over f_\pi}{m_\Xi-m_\Sigma\over 2m_s-m_u-m_d}(1-I),
\en
where
\be
I={m_\pi^2\over m_{\eta'}^2}\cos\phi\left(\cos\phi+{1\over\sqrt{2}}\sin\phi\right)-{m_\pi^2\over m_{\eta}^2}\sin\phi\left({1\over\sqrt{2}}\cos\phi-\sin\phi\right)
\en
comes from the $\eta_0$ tadpole contribution
and $\phi$ is the $\eta-\eta'$ mixing angle defined by
\be
\eta=\eta_8\cos\phi-\eta_0\sin\phi, \qquad \eta'=\eta_8\sin\phi+\eta_0\cos\phi.
\en
The expression of $\bar g_{\pi NN}$ in Eq. (\ref{eq:gpiNN}) without the correction $I$ is precisely the current-algebra result obtained in \cite{CDVW}.
In the limit $a\to 0$, one has
\be
m_\eta^2=2m_K^2-m_\pi^2, \quad m_{\eta'}^2=m_\pi^2, \quad \phi={\rm arctan}(1/\sqrt{2}).
\en
Hence, the $\eta'$ is as light as the pion in the absence of the axial anomaly. This is the well-known $U_A(1)$ problem. Since $I\to 1$ in the $a\to 0$ limit, it is evident that $\bar g_{\pi NN}$ vanishes when the gluonic anomaly is turned off, as it should be. Therefore, although the $\eta_0$ pole contribution is very small numerically, it ensures that the physical result has the correct behavior in chiral and vanishing anomaly limit.

Now returning back to Eq. (\ref{eq:L'}), it is clear that the \cp-odd operator $\delta{\cal L'}_{CP}^M$ does not have the right behavior in the $a\to 0$ limit, though it is perfect in the real world where $a$ is large. Indeed, the relation $a\bar\theta/N_c=2\bar mv\theta$ does not hold for $a\to 0$ at first place. This indicates that it is better to use either $\dL_{CP}^M$ or $\dL_{CP}^{\rm DVW}$ rather than  the hybrid one $\delta{\cal L'}_{CP}^M$.

\vskip 0.3cm {\bf 5.}  From $\dL_{CP}^M$ or $\dL_{CP}^{\rm DVW}$ it is easily seen that  strong \CP violation can manifest in the meson sector only if the number of the involved pseudoscalar mesons is odd. Therefore, strong \cp-violating effects at low energies can only be seen in the decays such as $(\eta,\eta',G)\to 2\pi, 4\pi$, $G\to K\bar K,\eta\eta$, and $(\eta,\eta',G)\to\gamma\gamma$ via $FF$ coupling with $G$ being a pseudoscalar glueball. Unfortunately, strong \cp-odd effects at low energies are always proportional to $\theta^2$ and hence extremely small except the baryon's electric dipole moment which is proportional to $\theta$. This is why the neutron electric dipole moment provides the most stringent limit on $\theta$. Nevertheless, $P$- and \cp-violating mestable
domains could be formed in heavy ion collisions at RHIC owing to the presence of the extremely strong magnetic field produced in such a collision (see e.g. \cite{RHIC} for a recent experiment). This provides a very exciting avenue to probe strong \cp-violating effects induced by the $\theta$ vacuum.

\vskip 0.3cm {\bf 6.} Whether
the strong \cp-violating Lagrangians (\ref{eq:LDV}) and $\dL_{CP}^{\rm DVW}$ obtained by Di Vecchia, Veneziano and Witten 3 decades ago are equivalent to the one $\dL_{CP}^{\rm Baluni}$ originally derived by Baluni at the quark or hadron level is studied in this work. A major difference between the two approaches is that the sum of the phases of the quark condensate is taken to be the same as the $\theta$ parameter (i.e. $\sum\phi_i=\theta$) in the latter,  while there is  a small and calculable deviation of $\sum\phi_i$ from $\theta$ in the former. As a consequence,  the $\theta G\tilde G$ term in the DVW approach is not entirely removed away after the vacuum is rotated from the \cp-odd state to the \cp-even one; strong \CP violation resides not only in the quark mass terms but also in the residual topological sector. Neglecting the $\eta_0$ tadpole contribution which is generally very small numerically, the DVW and Baluni Lagrangians are equivalent.
Contrary to some claims, it is necessary to include the SU(3)-singlet $\eta_0$ tadpole contribution for strong \cp-odd effects induced by the Baluni-type Lagrangian to ensure that strong \CP violation vanishes in the zero axial anomaly limit.

\vskip 1.71cm {\bf Acknowledgments}

We wish to thank the Physics Department of Brookhaven National
Laboratory and C.N. Yang
Institute for Theoretical Physics at SUNY Stony Brook for the hospitality. This research was supported in part by the National
Science Council of R.O.C. under Grant No. NSC97-2112-M-001-004-MY3.

\end{document}